\documentclass[nature-comm,superscriptaddress,floatfix,tightenlines,showpacs,twocolumn]{revtex4-1}
\usepackage{graphicx}
\usepackage{dcolumn}   
\usepackage{bm}        
\usepackage{amssymb}   
\usepackage{amsmath}
\usepackage{url}
\usepackage{layout}
\usepackage{color}
\usepackage{epsfig}
\usepackage{graphicx}
\usepackage{mathrsfs}\usepackage{bm}\usepackage{appendix}\usepackage{color}

\usepackage[thinlines]{easytable}
\usepackage{makecell}
\usepackage{tikz,pgf}
\usepackage{booktabs}
\usepackage{float}
\usepackage{hyperref}
\hypersetup{colorlinks,allcolors=blue}

\begin{document}

\title{Pairing mechanism and superconductivity in pressurized La$_5$Ni$_3$O$_{11}$}

\author{Ming Zhang}
\thanks{These two authors contributed equally to this work.}
\affiliation{Zhejiang Key Laboratory of Quantum State Control and Optical Field Manipulation,
Department of Physics, Zhejiang Sci-Tech University, 310018 Hangzhou, China}

\author{Cui-Qun Chen}
\thanks{These two authors contributed equally to this work.}
\affiliation{Center for Neutron Science and Technology, Guangdong Provincial Key Laboratory of Magnetoelectric Physics and Devices, State Key Laboratory of Optoelectronic Materials and Technologies, School of Physics, Sun Yat-Sen University, Guangzhou 510275, China}

\author{Dao-Xin Yao}
\email{yaodaox@mail.sysu.edu.cn}
\affiliation{Center for Neutron Science and Technology, Guangdong Provincial Key Laboratory of Magnetoelectric Physics and Devices, State Key Laboratory of Optoelectronic Materials and Technologies, School of Physics, Sun Yat-Sen University, Guangzhou 510275, China}

\author{Fan Yang}
\email{yangfan\_blg@bit.edu.cn}
\affiliation{School of Physics, Beijing Institute of Technology, Beijing 100081, China}


\begin{abstract}
\noindent The discovery of superconductivity (SC) with critical temperature $T_c$ above the boiling point of liquid nitrogen in pressurized La$_3$Ni$_2$O$_{7}$ has sparked a surge of exploration of high-$T_c$ superconductors in the Ruddlesden-Popper (RP) phase nickelates. More recently, the RP phase nickelate La$_5$Ni$_3$O$_{11}$, which hosts a layered structure with alternating bilayer and single-layer NiO$_2$ planes, has been reported to accommodate SC under pressure, exhibiting a dome-shaped pressure dependence with highest $T_c\approx 64$ K, capturing a lot of interest. Here, using density functional theory (DFT) and random phase approximation (RPA) calculations, we systematically study the electronic properties and superconducting mechanism of this material. Our DFT calculations yield a band structure including two nearly decoupled sets of sub-band structures, with one set originating from the bilayer subsystem and the other from the single-layer one. RPA-based analysis demonstrates that SC in this material occurs primarily within the bilayer subsystem exhibiting an $s^\pm$ wave pairing symmetry similar to that observed in pressurized La$_3$Ni$_2$O$_{7}$, while the single-layer subsystem mainly serves as a bridge facilitating the inter-bilayer phase coherence through the interlayer Josephson coupling (IJC). Since the IJC thus attained is extremely weak, it experiences a prominent enhancement under pressure, leading to the increase of the bulk $T_c$ with pressure initially. When the pressure is high enough, the $T_c$ gradually decreases due to the reduced density of states on the $\gamma$-pocket. In this way, the dome-shaped pressure dependence of $T_c$ observed experimentally is naturally understood.

\end{abstract}\maketitle

~~~~~~~~~~~~~~~~~~~~~~~~~~~~~~~~~~~~~~~~~~~~~~~~~~~~~~~~~~~~~~~~~~~~~~~~~~~~~~~~~~~~~~~~~~~~~~~~~~~~~~~~~~~~~~~~~~~~~~~~~~~~~~~~~~~~~~~~~~~~~~~

\noindent The discovery of superconductivity (SC) with critical temperature $T_c\approx 80$ K in the Ruddlesden-Popper (RP) phase bilayer nickelate La$_3$Ni$_2$O$_{7}$ under pressure~\cite{bilayernature} has sparked significant experimental and theoretical interest in the field~{\color{red}\cite{ bilayermodel,huomodulation,zhang2023electronic,lechermann2023electronic,luo2024high,wu2024superexchange,  cpl_41_7_077402,shilenko2023correlated,yang2023interlayer, zhang2023trends,PhysRevLett.131.206501,shen2023effective,Gu_2025,oh2023type,YangF2023,liao2023electron,WangQH2023,PhysRevB.110.235155,kaneko2024pair,ouyang2024absence,heier2024competing,zhang2024structural,zhang2024electronic,tian2024correlation,ryee2024quenched,zhang2024strong, ni_spin_2025,lu2024interlayer,qu2024bilayer,yang2024strong, fan2024superconductivity,   sakakibara2024possible, cao2024flat, jiang2024pressure,chen2025charge,yang_orbital-dependent_2024, NPzhang,Hou_2023, PhysRevB.110.134520, wang_pressure-induced_2024, liu_electronic_2024,PhysRevB.110.L180501,PhysRevLett.134.076001, shao2024possiblehightemperaturesuperconductivitydriven}}. Subsequently, researchers observed SC in the trilayer nickelate La$_4$Ni$_3$O$_{10}$ under pressure, and recently detected SC in ambient-pressure (AP) La$_3$Ni$_2$O$_{7}$ thin films~\cite{zhu_superconductivity_2024,zhang2025superconductivity,ko_signatures_2025,zhou_ambient-pressure_2025}. These findings further underscore the potential of RP-phase nickelates as high-$T_c$ superconductors~\cite{PhysRevB.110.014503, PhysRevB.111.075140,PhysRevB.109.144511, li_signature_2024,hu2025electronicstructuresmultiorbitalmodels,experimental,wang2024electronic,RN11}. In RP phase nickelates with the general formula R$_{n+1}$Ni$_n$O$_{3n+1}$ (R = rare earth), each Ni forms a NiO$_6$ octahedron with corner-sharing O atoms. At AP, bilayer La$_3$Ni$_2$O$_{7}$ and trilayer La$_4$Ni$_3$O$_{10}$ manifest octahedral tilting and in-plane lattice anisotropy~\cite{bilayernature,experimental,zhu_superconductivity_2024}. Upon increasing pressure, both the octahedral tilting and the in-plane lattice constant disparity gradually diminish and eventually the crystal structures attain higher symmetry ($I4/mmm$ space group)~\cite{li_identification_2025}.

Furthermore, hybrid RP phase nickelates, formed by alternating stacking different RP phases along the c-axis, also attract widespread investigations, especially the hybridization of the single-layer La$_2$NiO$_{4}$, the bilayer La$_3$Ni$_2$O$_{7}$ and the trilayer La$_4$Ni$_3$O$_{10}$~\cite{zhang2024magneticcorrelationspairingtendencies, chen_jpcm,PhysRevLett.133.146002, PhysRevMaterials.8.053401, ouyang2025phasediagramskeyfactors}. Until now, successful synthesis of 1313 phase (La$_3$Ni$_2$O$_{7}$) ~\cite{chen_jpcm,PhysRevLett.133.146002} and 1212 phase (La$_5$Ni$_3$O$_{11}$)~\cite{PhysRevMaterials.8.053401} has been achieved. A previous study reported a superconducting transition onset temperature up to 80 K in alternating stacking  single-layer La$_2$NiO$_{4}$ and trilayer La$_4$Ni$_3$O$_{10}$, namely 1313 phase~\cite{PhysRevLett.133.146002}. However, it remains unclear whether the observed high-$T_c$ comes from the hybrid 1313 phase or 2222 bilayer phase.

Recently, pressure-induced SC with maximal $T_c = $ 64 K was observed in the hybrid 1212 phase, i.e. La$_5$Ni$_3$O$_{11}$~\cite{shi2025superconductivityhybridruddlesdenpopperla5ni3o11}. Transport and magnetic torque measurements also revealed a density-wave (DW) transition at ambient and low pressures. In contrast to the previously reported DW in La$_3$Ni$_2$O$_{7}$ and La$_4$Ni$_3$O$_{10}$, the DW observed here is notably robust against pressure. Intriguingly, SC emerges at around 11.7 GPa and exhibits a dome-shaped pressure-dependent behavior, in contrast to the previously reported right-triangle-like behavior observed in pressurized La$_3$Ni$_2$O$_{7}$~\cite{li_identification_2025}. Previous random-phase approximation (RPA) based study proposed that $d$-wave is the leading pairing symmetry that originates from the La$_2$NiO$_{4}$ single-layer in the material~\cite{zhang2025electronicstructuremagneticpairing}. However, recent studies using combined density functional theory (DFT) and dynamic-mean-field-theory (DMFT) provide another possibility that the La$_2$NiO$_{4}$ single-layer in La$_5$Ni$_3$O$_{11}$ is nearly Mott-insulating which does not carry SC~\cite{ouyang2025phasediagramskeyfactors,labollita2025correlated}, consistent with previous insight that the pure La$_2$NiO$_{4}$ is an antiferromagnetic (AFM) Mott insulator~\cite{PhysRevB.75.012414,PhysRevB.80.144523}. Therefore, a more systematic investigation into the electronic and superconducting properties of the hybrid 1212 phase La$_5$Ni$_3$O$_{11}$ is urgently needed.

In this paper, we adopt the first-principle DFT calculations to study the electronic structure of La$_5$Ni$_3$O$_{11}$, followed by a RPA based study to clarify the superconducting mechanism in this material. Based on our DFT band structure, we construct a trilayer $(d_{z^2}, d_{x^2-y^2})$-orbital tight-binding (TB) model, in which the hopping integrals between the bilayer subsystem and the single-layer one is very weak, reflecting that the two subsystems are almost decoupled, allowing for the two parts to be treated separately. Our RPA based analysis suggests that the superconducting pairing mainly occurs in the bilayer subsystem, which takes the s$^\pm$-pattern similar with that in pressurized La$_3$Ni$_2$O$_{7}$, while the single-layer subsystem mainly serves as a bridge connecting adjacent superconducting bilayer subsystems to establish phase coherence along the c-axis necessary for the bulk SC, through the interlayer Josephson coupling (IJC). As the IJC is extremely weak, it will experience prominent enhancement with the increase of the pressure, leading to enhancement of the $T_c$ in the low pressure regime. When the pressure is strong enough, the $T_c$ decreases with further enhancement of the pressure due to a reduction in the density of states (DOS) on the $\gamma$-pocket crucial for pairing. Therefore, our study provides a natural understanding of the dome-shaped pressure dependence of the $T_c$ observed in the experiment. Furthermore, our findings elucidate the distinct pressure responses of hybrid-phase and pure-phase nickelate superconductors, offering a unified framework that may also be extended to other hybrid superconducting systems.


~~~~~~~~~~~~~~~~~~~~~~~~~~~~~~~~~~~~~~~~~~~~~~~~~~~~~~~~~~~~~~~~~~~~~~~~~~~~~~~~~~~~~~~~~~~~~~~~~~~~~~~~~~~~~~~~~~~~~~~~~~~~~~~~~~~~~~~~~~~~~~~

\noindent{\bf Results}


\noindent{\bf Band Structure and TB model} 

\begin{figure}[htbp]
\centering
\includegraphics[width=0.45\textwidth]{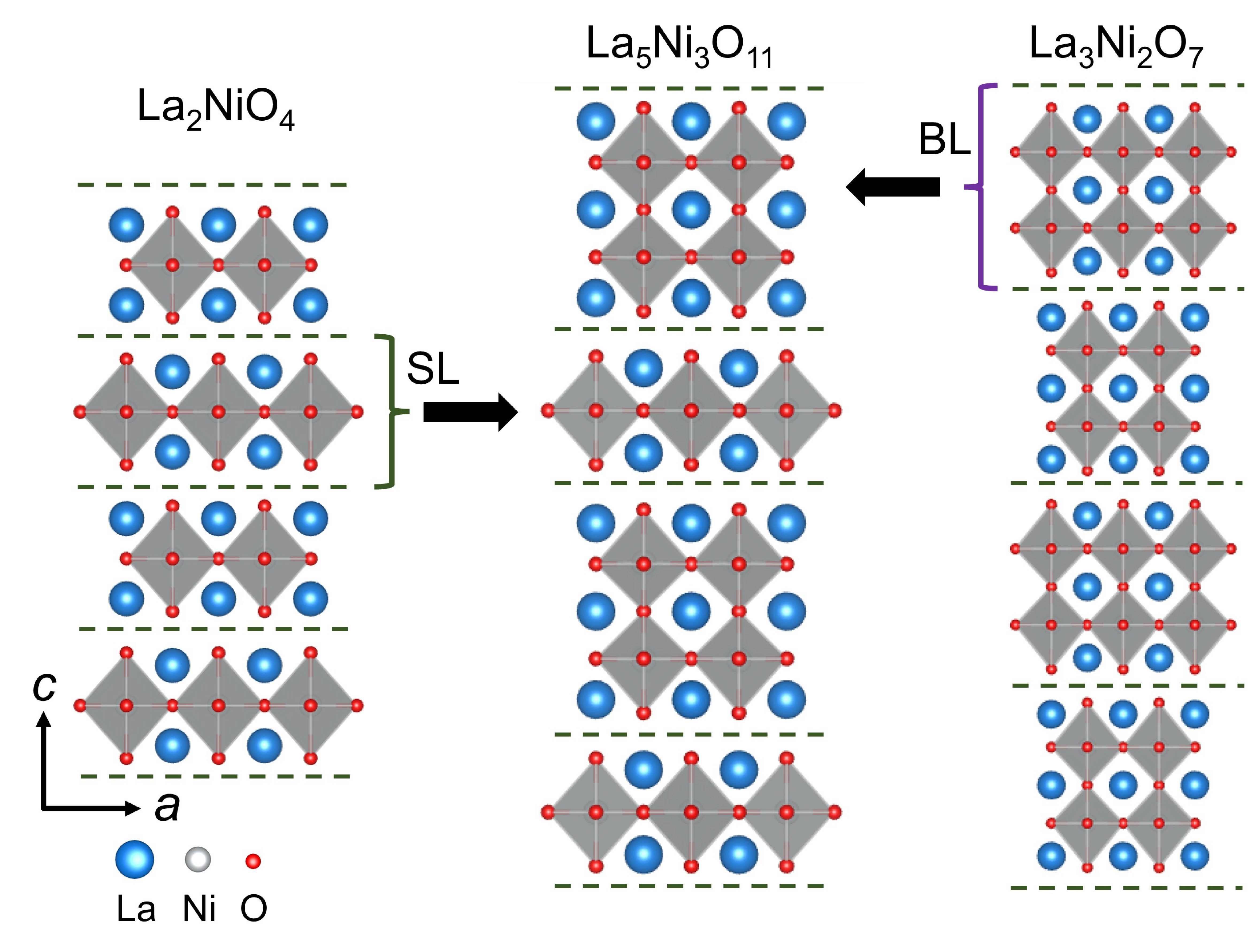}
\caption{(color online) Side view of crystal structures of the single-layer (La$_2$NiO$_{4}$), bilayer (La$_3$Ni$_2$O$_{7}$) and SL-BL (La$_5$Ni$_3$O$_{11}$) RP phase nickelates. The blue, grey, and red balls represent lanthanum, nickel, and oxygen atoms, respectively. }
\label{fig1}
\end{figure}

\noindent La$_5$Ni$_3$O$_{11}$ is a single-layer-bilayer (SL-BL) stacked nickelate, as shown in Fig.~\ref{fig1}. 
La$_2$NiO$_{4}$ and La$_3$Ni$_2$O$_{7}$ stack alternatively in the RP phase, crystallizing into an orthorhombic phase ($Cmmm$ or $Immm$ space group) at AP and transitioning to $P4/mmm$ space group under high pressure (HP)~\citep{shi2025superconductivityhybridruddlesdenpopperla5ni3o11,PhysRevMaterials.8.053401}. DFT calculated phonon spectra show imaginary frequencies at AP whereas the imaginary frequencies vanish under HP for $P4/mmm$ phase (see Fig.S1~\cite{SI}), demonstrating a structural phase transition under pressure and the structural stability of $P4/mmm$ phase under high pressure. Unlike bulk La$_3$Ni$_2$O$_{7}$, the out-of-plane Ni-O-Ni angle in  La$_5$Ni$_3$O$_{11}$ remains 180$^{\circ}$ from AP to HP, without octahedral rotation. Therefore, in La$_5$Ni$_3$O$_{11}$ each unit cell contains only one Ni atom per layer, in contrast to the cases of bulk La$_3$Ni$_2$O$_{7}$ and La$_4$Ni$_3$O$_{10}$ in which each unit cell contains two Ni atoms per layer at AP due to octahedral rotation.

\begin{figure}[htbp]
\noindent
\centering
\includegraphics[width=0.5\textwidth]{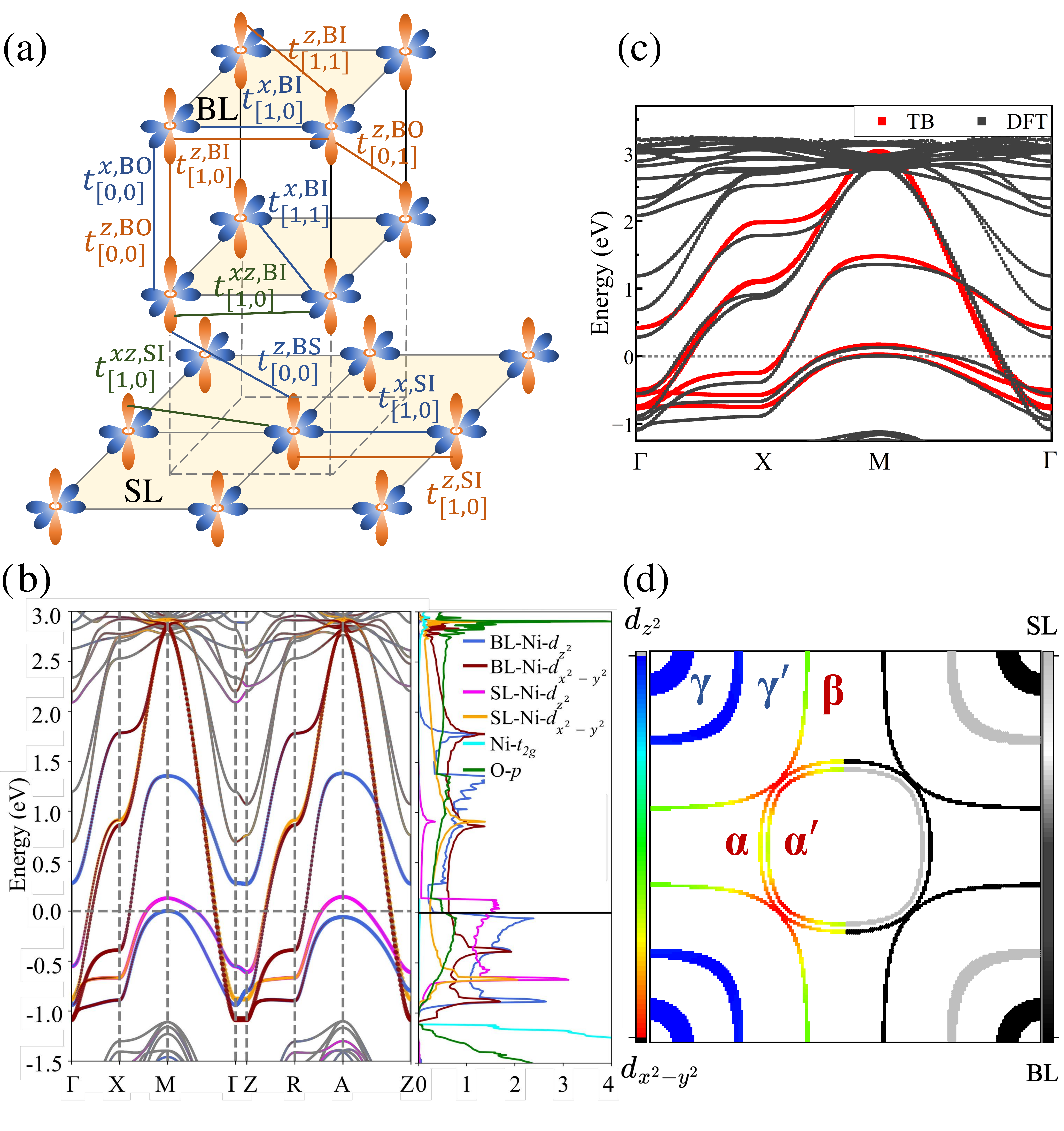}
\caption{(color online) Band structure of the DFT and six-orbital TB model for La$_5$Ni$_3$O$_{11}$ at 12 GPa. (a) DFT band structure and projected DOS of La$_5$Ni$_3$O$_{11}$. (b) FS in the BZ, with the five pockets labeled. The color scheme in the left half of (b) represents the relative contributions of the $d_{z^2}$ and $d_{x^2-y^2}$ orbital, while the colors scheme on the right half of (b) indicates the relative contributions from Ni atoms in the single-layer and bilayer subsystems.
(c) Schematic of La$_5$Ni$_3$O$_{11}$ lattice of six-orbital TB model. The dashed line denotes the $(\frac{1}{2},\frac{1}{2})$ translation of RP stacking between bilayer and single-layer sublattice. (d) The TB band structure corresponding to (c).}
\label{fig2}
\end{figure}

The band structure of the SL-BL stacking La$_5$Ni$_3$O$_{11}$ at 12 GPa is shown in Fig.~\ref{fig2}(a), consistent with previous theoretical results~\cite{ouyang2025phasediagramskeyfactors,labollita2025correlated}. As in most RP phase nickelates, bands around the Fermi level can be described by the Ni-$e_g$ sector. The SL-derived Ni-$d_{z^2}$ band lies between the BL-derived bonding and anti-bonding bands. Compared to the band structure at AP (see Fig.S2~\cite{SI}), the BL-derived bonding band is metalized by pressure, leaving the $\gamma$ pocket at the brillouin zone (BZ) corner, as shown in Fig.~\ref{fig2}(b). This phenomenon resembles the behavior of pressurized bulk La$_3$Ni$_2$O$_7$.

To further investigate the electronic properties in La$_5$Ni$_3$O$_{11}$, we perform Wannier downfolding on the Ni-$d_{z^2}$ and $d_{x^2-y^2}$ orbitals based on the DFT electronic structure. The obtained TB Hamiltonian in real space can be expressed as
\begin{align}
H_{{\rm TB}}
=\sum_{\bm r_i\Delta \bm r\mu\nu\sigma}t_{\Delta \bm r\mu\nu}
c^{\dagger}_{\bm r_i\mu\sigma}c_{(\bm r_i+\Delta \bm r)\nu\sigma}.
\label{eq1}
\end{align}
Here $\bm r_i$ represents the coordinates of site $i$, $\Delta \bm r_{x}(\Delta \bm r_{y})\in(-2,2)$ represents the hoppings up to the third-nearest neighbor. In this context, $\bm{r}_i$ refers to the unit cell index, while a finite $\Delta \bm{r}$ represents the interunit cell separation. The indices $\mu, \nu=1, \cdots, 6$ containing the $d_{z^2}$ and $d_{x^2-y^2}$ orbitals of the upper and lower layers of the bilayer subsystem, as well as those of the single-layer subsystem. The SL-BL lattice and corresponding hoppings are depicted in Fig.~\ref{fig2}(c), with the hopping integrals $t_{ij}^{\mu\nu}$ provided in Tab.~\ref{tab:hopping}. Note that due to the RP stacking in La$_5$Ni$_3$O$_{11}$, a $(\frac{1}{2},\frac{1}{2})$ translation exists between the bilayer and single-layer sublattice. Fig.~\ref{fig2}(d) displays the band structure for TB model, which is in good agreement with DFT results. The obtained FS for this TB model contains five pockets labeled as $\alpha$, $\alpha^\prime$, $\beta$, $\gamma$, $\gamma^\prime$, as shown in Fig.~\ref{fig2}(b). The dominant component orbital of $\gamma$ and $\gamma^\prime$ pockets is $d_{z^2}$, while the $\alpha$, $\alpha^\prime$ and $\beta$ pockets are the mix of $d_{z^2}$ and $d_{x^2-y^2}$ orbitals. From the right half of Fig.~\ref{fig2}(b), we can see that the $\alpha^\prime$ and $\gamma^\prime$ pocket are contributed from single-layer subsystem while the other three pockets from bilayer subsystem.

\begin{table}[t]
\renewcommand{\arraystretch}{1.5}
\caption{Hopping parameters and site energies of TB model for La$_5$Ni$_3$O$_{11}$ at 12 GPa in unit of eV. Here, $x$, $z$ and $xz$ denotes hopping within and between $d_{x^{2}-y^{2}}$, $d_{z^{2}}$ orbitals, respectively. The abbreviations BI, BO, and SI represent intra-bilayer, inter-bilayer, and single-layer in-plane hoppings, respectively, while BS denotes hoppings between bilayer and single-layer units. The TB model includes hopping processes up to the third-nearest neighbor, i.e., [2,0]. $\epsilon$ denotes the on-site energy.}
\noindent\begin{centering}
\begin{tabular}{ccccccc}
\hline \hline 
 $t_{[1,0]}^{z,BI}$ & $t_{[1,1]}^{z,BI}$& $t_{[1,0]}^{xz,BI}$ & $t_{[1,0]}^{x,BI}$ & $t_{[1,1]}^{x,BI}$ & $t_{[0,0]}^{z,BO}$  & $t_{[1,0]}^{z,BO}$\tabularnewline 
-0.1125 & -0.0180 & 0.2373 & -0.4708 & 0.0661 & -0.6713 & 0.0202 \tabularnewline  \hline
 $t_{[1,1]}^{z,BO}$ & $t_{[0,0]}^{x,BO}$ & $t_{[1,0]}^{x,BO}$ & $t_{[1,1]}^{x,BO}$ & $t_{[1,0]}^{xz,BO}$ & $t_{[\frac{1}{2},\frac{1}{2}]}^{z,BS}$ & $t_{[1,0]}^{z,SI}$ \tabularnewline 
 0.0065 & 0.0132& -0.0013 & 0.0017 & -0.0320 & -0.0103 & -0.0886 \tabularnewline \hline
 
  $t_{[1,1]}^{z,SI}$ & $t_{[1,0]}^{xz,SI}$ & $t_{[1,0]}^{x,SI}$ &  $t_{[1,1]}^{x,SI}$  & $t_{[2,0]}^{z,BI}$ & $t_{[2,0]}^{xz,BI}$ & $t_{[2,0]}^{x,BI}$  \tabularnewline  
  -0.0108 & 0.1901 & -0.4522 & 0.0752 & -0.0192&  0.0344& -0.0686 \tabularnewline  \hline

 $t_{[2,0]}^{z,SI}$ & $t_{[2,0]}^{xz,SI}$ & $t_{[2,0]}^{x,SI}$  & $\epsilon_z^B$  & $\epsilon_x^B$ & $\epsilon_z^S$ &  $\epsilon_x^S$ \tabularnewline  
 -0.0121 & 0.0256  & -0.0590 & 0.3288 & 0.8648 & 0.1360 & 0.9313  \tabularnewline 
\hline \hline 
\end{tabular}
\label{tab:hopping}
\par\end{centering}
\end{table}

~~~~~~~~~~~~~~~~~~~~~~~~~~~~~~~~~~~~~~~~~~~~~~~~~~~~~~~~~~~~~~~~~~~~~~~~~~~~~~

\begin{figure}[htbp]
\centering
\includegraphics[width=0.45\textwidth]{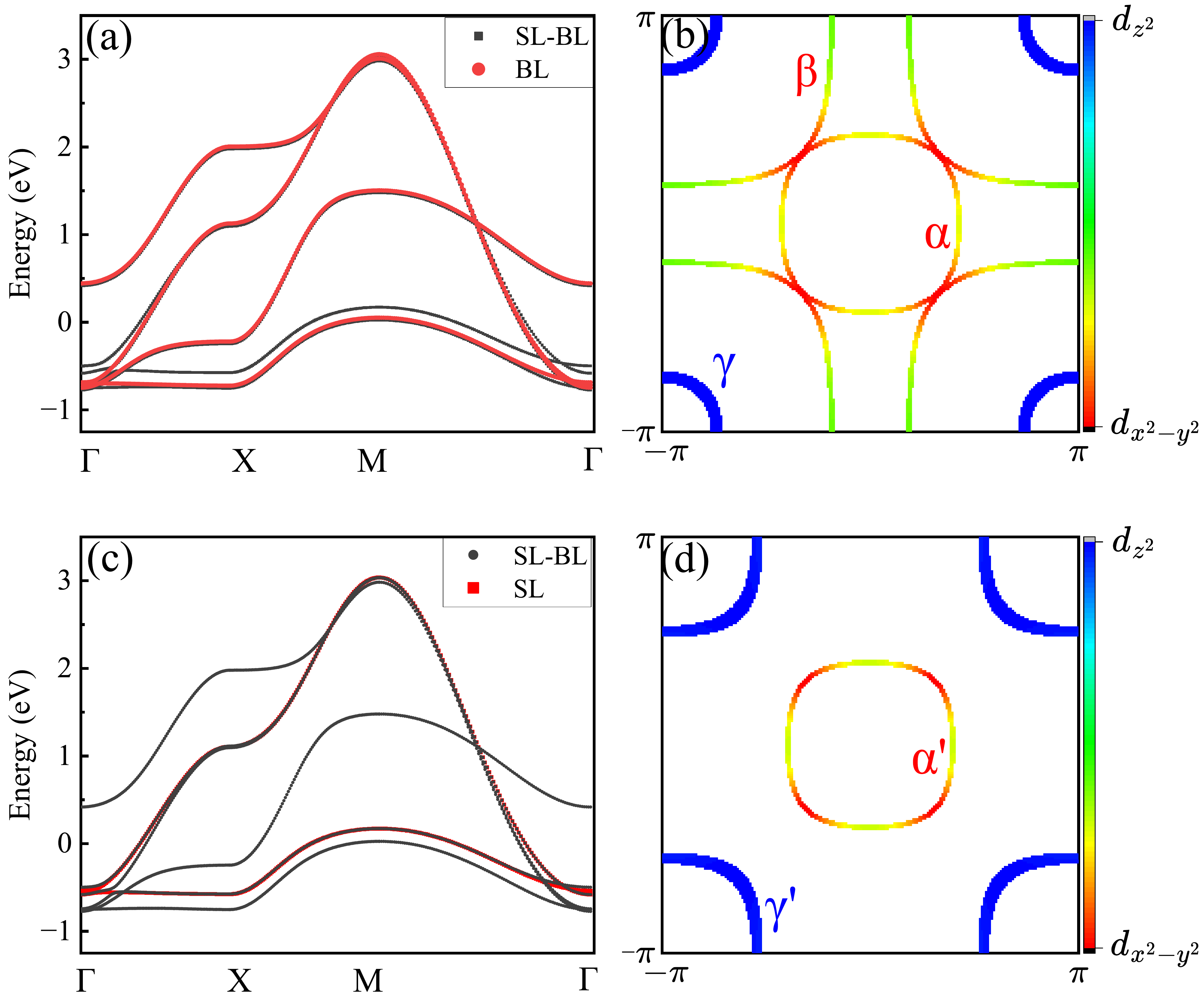}
\caption{(color online) TB bands and FS characteristics of the BL and SL subsystems in La$_5$Ni$_3$O$_{11}$ (SL-BL system) at 12 GPa. (a) Band structure of the SL-BL system (black lines) and that of the isolated BL subsystem (red lines). (b) FS of the isolated BL subsystem, showing that the $\alpha$ pocket is primarily composed of $d_{x^2-y^2}$ orbitals, the $\beta$ pocket contains contributions from both $d_{x^2-y^2}$ and $d_{z^2}$ orbitals, and the $\gamma$ pocket originates mainly from $d_{z^2}$ orbitals. (c) Band structure of the SL-BL system (black lines) and the isolated SL subsystem (red lines). (d) FS of the isolated SL subsystem, showing that the $\alpha'$ pocket is primarily composed of $d_{x^2-y^2}$ orbitals, the $\gamma'$ pocket originates mainly from $d_{z^2}$ orbitals.}
\label{fig3}
\end{figure}

    \noindent As shown in Tab.~\ref{tab:hopping}, the coupling between the single-layer subsystem and the bilayer subsystem is extremely weak, with the strongest hopping integral amplitude between the two parts being $\left|t_{[\frac{1}{2},\frac{1}{2}]}^{z,BS}\right|$=0.0103 eV.  The sole effect of such a weak coupling is to equilibrate the chemical potentials of the two subsystems and thereby adjust the electron filling. Consequently, it is reasonable to treat the two subsystems separately. 

The decoupling of the two subsystems at the TB Hamiltonian level is achieved by modifying the summation over $\mu$ and $\nu$ in Eq.~\ref{eq1}. Specifically, restricting $\mu, \nu \in (1, \cdots, 4)$ and electron filling $n\approx3.1$ which got from SL-BL system yields the Hamiltonian of the bilayer subsystem, while restricting $\mu, \nu \in (5, 6)$ and electron filling $n\approx1.9$ yields that of the single-layer subsystem. A comparison between the band structure of the bilayer subsystem and that of La$_5$Ni$_3$O$_{11}$ (SL-BL system) is shown in Fig.~\ref{fig3}(a), where it can be seen that the bilayer subsystem reproduces the key features of the bilayer-derived bands in the full system. The three FS pockets of the bilayer subsystem, labeled $\alpha$, $\beta$, and $\gamma$, are shown in Fig.~\ref{fig3}(b), and are clearly separated from the five pockets observed in Fig.~\ref{fig2}(b). Fig.~\ref{fig3}(c) shows that the band structure of the single-layer subsystem matches well with the single-layer-derived bands of the full SL-BL system. The two Fermi pockets of the single-layer subsystem, corresponding to the $\alpha'$ and $\gamma'$ pockets of the full system, are shown in Fig.~\ref{fig3}(d).


\begin{figure}[htbp]
\centering
\includegraphics[width=0.45\textwidth]{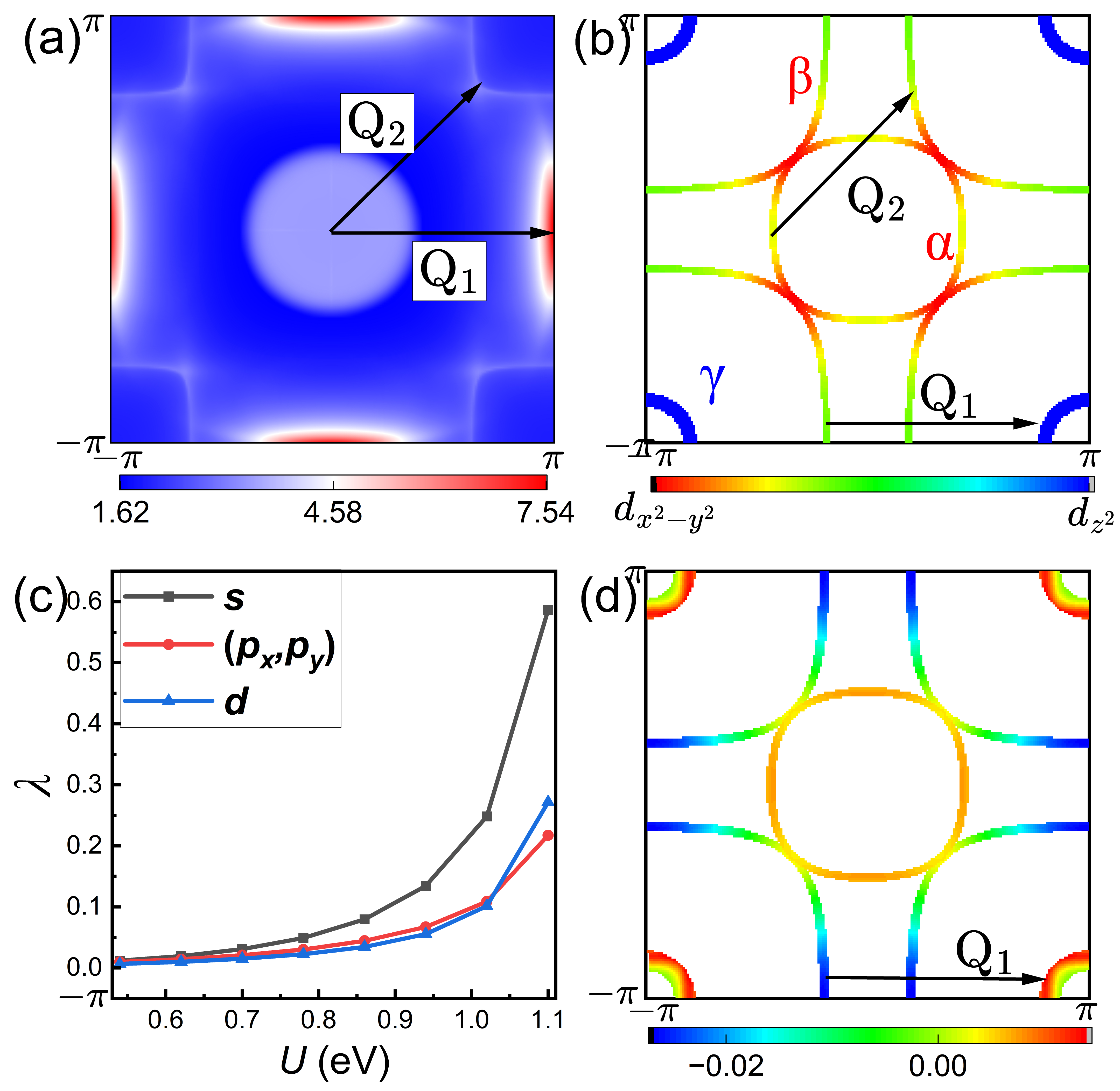}
\caption{(color online) (a) Distribution of the largest eigenvalue of the spin susceptibility matrix $\chi^{s}(q)$ in the BZ in the bilayer subsystem for $U = 1$ eV and $J_H = U/6$. The susceptibility peaks at two inequivalent momenta, denoted as Q$_1$ and Q$_2$, respectively. (b) FS of the bilayer subsystem in the BZ at 12 GPa. As shown in (b), Q$_1$ corresponds to a nesting vector between the $\beta$ and $\gamma$ pockets, while Q$_2$ corresponds to a nesting vector between the $\alpha$ and $\beta$ pockets. (c) The largest pairing eigenvalue $\lambda$ of the various
pairing symmetries as function of the interaction strength $U$ with
fixed $J_H= U/6$. (d) The distributions of the leading $s$-wave pairing gap function on the FS for $U=1.1$ eV, exhibiting an $s^{\pm}$-pattern.}
\label{fig4}
\end{figure}

The separation of the single-layer and bilayer subsystems within the TB framework points toward an underlying layer-selective character in this hybrid RP-phase nickelate. At ambient and low pressures, the incorporation of local dynamical correlations via the DMFT modifies the DFT band structure by removing the FS pockets contributed by the single‐layer subsystem, leaving only the three FS pockets contributed by the bilayer subsystem~\cite{ouyang2025phasediagramskeyfactors,labollita2025correlated}. This observation indicates that the single-layer subsystem is close to a Mott‐insulating state, with possible AFM order present~\cite{ouyang2025phasediagramskeyfactors}, analogous to the bulk La$_2$NiO$_4$~\cite{PhysRevB.75.012414,PhysRevB.80.144523}. At HP, the Mott gap becomes narrower, permitting a small density of mobile carriers that may suppress the AFM order~\cite{ouyang2025phasediagramskeyfactors}. Indeed, the long‐range magnetic order has not been observed experimentally under HP so far~\cite{PhysRevMaterials.8.053401}. Regardless of whether residual magnetic order persists or not, the single-layer subsystem remains a ``bad metal'' and can hardly carry SC~\cite{ouyang2025phasediagramskeyfactors}. This suggests that superconductivity is more likely to originate from the bilayer subsystem. Since DFT+DMFT indicates that electronic correlation in the bilayer subsystem is comparatively weaker, we will focus our subsequent RPA analysis on the bilayer subsystem to study its pairing instability.

~~~~~~~~~~~~~~~~~~~~~~~~~~~~~~~~~~~~~~~~~~~

\noindent{\bf RPA study of the SC}

Given that superconductivity is likely to arise predominantly from the bilayer subsystem, we adopt the following multi-orbital Hubbard interaction to investigate the SC driven by electron interactions in this bilayer subsystem,
\begin{align}\label{hubbard}
H_{int}&=U\sum_{i\tilde{\mu}}n_{i\tilde{\mu}\uparrow}n_{i\tilde{\mu}\downarrow}+
V\sum_{i,\sigma,\sigma^{\prime}}n_{i1\sigma}n_{i2\sigma^{\prime}} \nonumber\\
&+J_{H}\sum_{i\sigma\sigma^{\prime}} \Big[c^{\dagger}_{i1\sigma}c^{\dagger}_{i2\sigma^{\prime}}c_{i1\sigma^{\prime}}c_{i2\sigma}+(c^{\dagger}_{i1\uparrow}c^{\dagger}_{i1\downarrow}c_{i2\downarrow}c_{i2\uparrow}+h.c.)\Big].
\end{align}
Here, $U$, $V$, and $J_H$ denote the intra-orbital, inter-orbital Hubbard repulsion, and the coupling of Hund (and the pair hopping) respectively, which satisfy the relation $U=V+2J_H$. $\tilde{\mu} \in (1,2)$ labels the two orbitals ($d_{z^2}$ and $d_{x^2-y^2}$) associated with each Ni atom. $i$ denotes the lattice sites belonging to the bilayer subsystem. We employ the multi-orbital RPA approach ~\cite{takimoto2004strong,yada2005origin,kubo2007pairing,graser2009near,liu2013d+,zhang2022lifshitz,kuroki101unconventional} to treat this Hamiltonian. By renormalization, the spin susceptibility $\bm{\chi}^{(s)}$ and charge susceptibility $\bm{\chi}^{(c)}$ can be defined as Eq. \eqref{chisce}:
\begin{align}\label{chisce}
 \bm{\chi}^{(s)}(\bm {k},i\nu)=[I-\bm{\chi}^{(0)}(\bm {k},i\nu)
 U^{(s)}]^{-1}\bm{\chi}^{(0)}(\bm {k},i\nu),\nonumber\\
 \bm{\chi}^{(c)}(\bm {k},i\nu)=[I+\bm{\chi}^{(0)}(\bm {k},i\nu)
 U^{(c)}]^{-1}\bm{\chi}^{(0)}(\bm {k},i\nu).
\end{align}
 Where $\bm{\chi}^{(0)}$ is bare susceptibility for the non-interacting case, expressed as a tensor $\bm{\chi}^{(0)pq}_{st}$, with $p,q,s,t$ as orbital indices. Similarly, $\bm{\chi}^{(s)}$ can be expressed as $\bm{\chi}^{(s)pq}_{st}$. $U^{(s,c)}$ is the renormalized interaction strength, which is represented as a $4^2\times 4^2$ matrix in the bilayer subsystem. Note that there is a critical interaction strength $U_c^{(s/c)}$ for the spin/charge susceptibility. When $U\geq U_c^{(s,c)}$, the denominator matrix in Eq. \eqref{chisce} will have zero eigenvalues for certain values of $\bm{k}$, causing the renormalized spin or charge susceptibility to diverge. This divergence indicates the onset of magnetic or charge order. Generally, repulsive Hubbard interactions suppress the charge susceptibility, but enhance the spin susceptibility ~\cite{takimoto2004strong,yada2005origin,kubo2007pairing,graser2009near,liu2013d+,zhang2022lifshitz,kuroki101unconventional}. Therefore, $U_c^{(s)}<U_c^{(c)}$ and we denote $U_c^{(s)}$ as $U_c$ in the bilayer subsystem in the following. When fixing $J_H = U/6$, we find  $U_c\approx 1.15$ eV.

Defining $\chi^s(\bm {q})$ as the maximum eigenvalue of $\bm{\chi}^{(s)}$ at the momentum $\bm {q}$, Fig.~\ref{fig4}(a) shows its distribution in the BZ  for $U=1$ eV$<U_c$ of the bilayer subsystem. Notably, the distribution exhibits peaks at two unequivalent momenta, which we have labeled as $\mathbf{Q}_1$ and $\mathbf{Q}_2$. These two momenta precisely correspond to the two FS nesting vectors of the bilayer subsystem, as illustrated in Fig.~\ref{fig4}(b), and the vector $\mathbf{Q}_1=(\pi,0)$ is associated with the highest intensity of spin susceptibility, which is very similar to the case of bulk La$_3$Ni$_2$O$_7$ under pressure~\cite{WangQH2023,YangF2023}. 

When $U<U_c$, the spin fluctuations can mediate SC in the bilayer subsystem, whose $T_c$ is related to the largest pairing eigenvalue $\lambda$ via $T_c\propto \omega_De^{-1/\lambda}$~\cite{graser2009near}, where the ``Debye frequency'' $\omega_D$ represents the typical energy scale of spin fluctuations. The pairing symmetry is determined by the corresponding eigenvector (see $\bf Methods$). Fig.~\ref{fig4}(c) illustrates the dependence of the largest pairing eigenvalue, denoted as $\lambda$, on the interaction strength $U$ for different potential pairing symmetries. The $D_{4h}$ point group of this bilayer subsystem structure allows for several possible pairing symmetries. In Fig.~\ref{fig4}(c), we show the three leading pairing symmetries: $s$-wave, $d$-wave, and degenerate $(p_x, p_y)$-wave pairings. It is clear that the $s$-wave is the leading pairing symmetry and dominates other ones. The gap function of the obtained $s$-wave pairing is shown on the FS in Fig.~\ref{fig4}(d), which displays the $s^{\pm}$ pattern. Consequently, the $\alpha$- and $\gamma$- pockets connected by the nesting vector $\mathbf{Q}_1$ are distributed with the strongest pairing amplitude, with opposite gap signs. This pairing pattern is also similar to that in pressurized  bulk La$_3$Ni$_2$O$_7$~\cite{luo2024high,WangQH2023,YangF2023}.

~~~~~~~~~~~~~~~~~~~~~~~~~~~~~~~~~~~~~~~~~~~~~~~~~~~~~~~~~~~~~~~~~~~~~~~~~~~~~~~~~~~~~~~~~~~~~~~~~~~~~~~~~~~~~~~~~~~~~~~~~~~~~~~~~~~~~~~~~~~~~~~

\noindent{\bf Pressure Dependence of the SC}

\begin{figure}[htbp]
\centering
\includegraphics[width=0.45\textwidth]{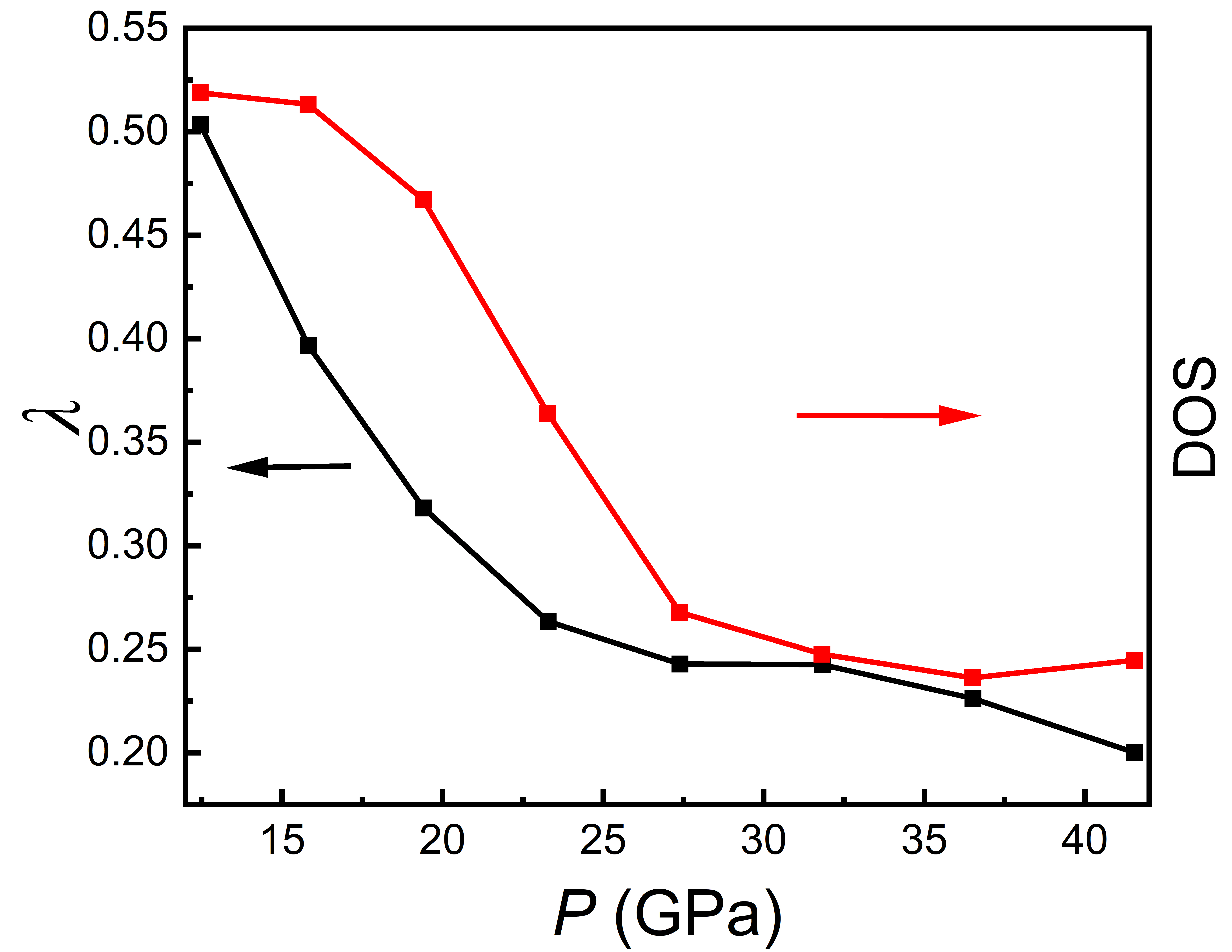}
\caption{(color online) Pressure-dependence of the pairing eigen value $\lambda$ (black solid line) and the DOS on the $\gamma$-pocket (red solid line).}
\label{fig5}
\end{figure}

\begin{table*}[htbp]
	\caption{Hopping parameters and site energies of TB model for La$_5$Ni$_3$O$_{11}$ under various pressures. $x$, $z$ and $xz$ denotes hopping within and between $d_{x^{2}-y^{2}}$, $d_{z^{2}}$ orbitals, respectively. The abbreviations BI, BO, and SI represent intrabilayer, interbilayer, and single-layer in-plane hoppings, respectively. Only hopping integrals greater than 1 meV are listed here for clarity. All parameters are in unit of eV. }
	\centering
	\setlength{\tabcolsep}{7.3pt}
	\renewcommand{\arraystretch}{1.5}
	\begin{tabular}{ccccccccccc}
		\hline
		\hline
		Pressure (GPa) & $\varepsilon_z^B$ & $\varepsilon_x^B$ & $\varepsilon_z^S$ & $\varepsilon_x^S$ & $t_{[0,0]}^{z,BO}$ & $t_{[1,0]}^{x,BI}$ &$t_{[1,0]}^{z,BI}$ & $t_{[1,0]}^{xz,BI}$  & $t_{[1,0]}^{x,SI}$  & $t_{[1,0]}^{xz,SI}$ \\ \hline
   
12 & 0.334 & 0.915 & -0.162 & 1.022  & -0.673  & -0.477   & -0.113  & 0.241  & -0.460  & 0.187  \\ 
16 & 0.335 & 0.921 & -0.163 & 1.024  & -0.695  & -0.487   & -0.122  & 0.245  & -0.469  & 0.192  \\ 
19 & 0.337 & 0.922 & -0.185 & 1.027  & -0.724  & -0.501   & -0.122  & 0.251  & -0.478  & 0.196  \\ 
23 & 0.337 & 0.930 & -0.188 & 1.029  & -0.749  & -0.511   & -0.128  & 0.256  & -0.489  & 0.1200 \\ 
27 & 0.338 & 0.931 & -0.207 & 1.029  & -0.759  & -0.523   & -0.131  & 0.261  & -0.500  & 0.203  \\ 
32 & 0.338 & 0.941 & -0.210 & 1.032  & -0.782  & -0.534   & -0.136  & 0.266  & -0.510  & 0.208  \\ 
37 & 0.339 & 0.942 & -0.221 & 1.036  & -0.801  & -0.546   & -0.142  & 0.271  & -0.523  & 0.212 \\  
42 & 0.340 & 0.951 & -0.234 & 1.035  & -0.812  & -0.557  & -0.146  & 0.275  & -0.534  & 0.216  \\ 

		\hline
		\hline
	\end{tabular}
    \label{tab2}
\end{table*}

To investigate the pressure dependence of SC in the system, we first performed DFT calculations of the electronic band structure under pressures ranging from approximately 12 to 40 GPa. The corresponding key hopping parameters at each pressure are listed in Tab.~\ref{tab2}. Based on these TB results, we employed the RPA approach to calculate the pairing eigenvalue $\lambda$ as a function of pressure $P$. As shown in Fig.~\ref{fig5}, the resulting $\lambda$ exhibits a downward trend with enhanced $P$, which originates from reduced DOS on the $\gamma$ pocket. 

The decreasing $\lambda\sim P$ relation caused by the reduced DOS shown in Fig.~\ref{fig5} parallels that in the bulk La$_3$Ni$_2$O$_7$~\cite{PhysRevLett.134.076001,huomodulation,zhang2024structural}, which closely matches the experimentally observed $T_c\sim P$ relation in that material~\cite{li_identification_2025}.  However, it seems at a glance that such a decreasing $\lambda\sim P$ relation conflicts with the $T_c\sim P$ relation in La$_5$Ni$_3$O$_{11}$, which shows a pronounced increase up to around 20 GPa, followed by a gradual decline afterward~\cite{shi2025superconductivityhybridruddlesdenpopperla5ni3o11}. The discrepancy between the $\lambda\sim P$ and the $T_c\sim P$ relations in La$_5$Ni$_3$O$_{11}$ originates from that its alternating SL-BL structure brings into extremely weak IJC, which is crucial for establishing bulk superconductivity. Since superconducting pairing originates from the bilayer subsystem while global phase coherence relies on interlayer coupling, the enhancement of IJC under pressure becomes decisive, thereby resolving the seeming contradiction. In the following, we clarify this viewpoint. 


\begin{figure}[htbp]
\centering
\includegraphics[width=0.45\textwidth]{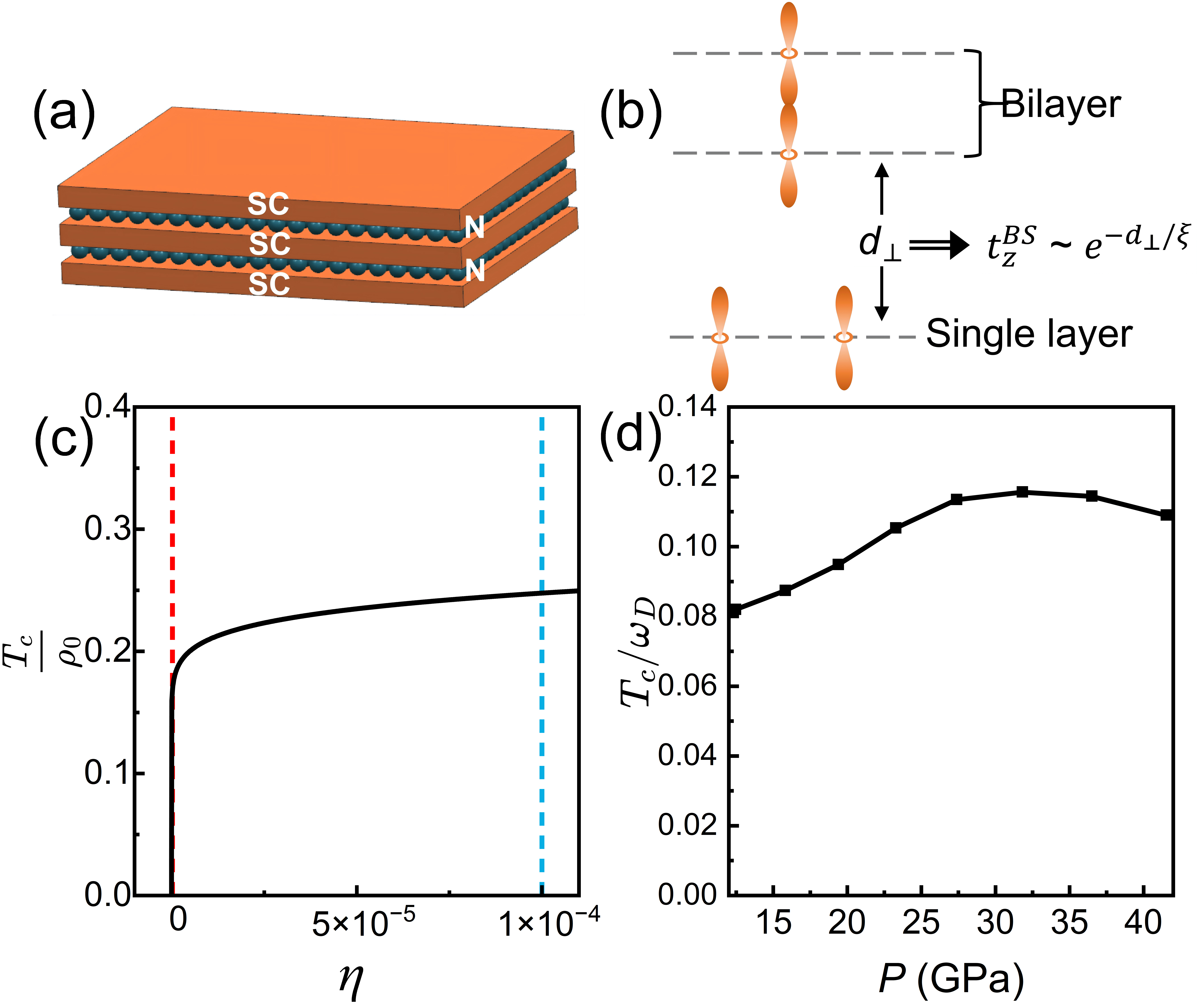}
\caption{(color online) Influence of the IJC on the pressure-dependence of the $T_c$. (a) Schematic of the intrinsic S–N–S Josephson Junction structure of La$_5$Ni$_3$O$_{11}$, where the superconducting (S) layers correspond to the BL subsystem and the normal (N) layers correspond to the SL subsystem. (b) Schematic of the interlayer distance $d_\perp$ dependence of the hopping integral $t_{z}^{BS}$ between the BL and the SL subsystems. (c) $T_c/\rho_0$ as a function of the anisotropy parameter $\eta$ that reflects the strength of the IJC. The red dashed line marks the $\eta\approx \mathrm 10^{-8}$ associated with the La$_5$Ni$_3$O$_{11}$, while the blue dashed line marks the $\eta\approx\mathrm 10^{-4}$ associated with the bulk La$_3$Ni$_2$O$_7$. (d) Pressure-dependence of the superconducting $T_c$ (in unit of $\omega_D$) in La$_5$Ni$_3$O$_{11}$.}
\label{fig6}
\end{figure}

As shown in Fig.~\ref{fig6}(a), the La$_5$Ni$_3$O$_{11}$ hosts alternating bilayer and single-layer NiO$_2$ subsystems, where the former supports SC and the latter is a ``bad metal". This results in an intrinsic superconductor–normal metal–superconductor (S–N–S) Josephson Junction structure. In this structure, the establishment of bulk SC requires not only intra-bilayer pairing but also inter-bilayer phase coherence. Given that the intermediate single-layer subsystem is close to a Mott-insulating state, it may be approximated as an intermediate layer without itinerant electrons participating in superconductivity. Physically, the inter-bilayer phase coherence is established via combined tunneling of a Cooper pair from a superconducting bilayer to its adjacent one, i.e. the IJC, across the intervening normal-metallic single-layer. This is a second-order process with respect to the tiny inter-bilayer single-particle tunneling, and hence a fourth-order process with respect to the weak bilayer-single-layer tunneling for a single particle. As such attained IJC is extremely weak, it will experience a prominent enhancement with the increase of the pressure. In the following, we provide a more quantitative clarification for this point.

In principle, the hopping integral $t_z^{BS}$ between the bilayer and the single-layer subsystems can be obtained through fitting the TB model to the DFT band structure. However, due to the limited accuracy achieved in the fitting, the obtained tiny $t_z^{BS}$ bears a considerably large relative error bar, and the accurate pressure dependence of $t_z^{BS}$ can be hardly acquired. To overcome this difficulty, we adopt the following semi-phenomenological analysis for the problem. As shown in Fig.~\ref{fig6}(b), the $t_z^{BS}$ depends on the overlap between the $d_{z^2}$ orbital wavefunctions of the bilayer and the single-layer subsystems, which decays exponentially with the displayed interlayer distance $d_{\perp}$, i.e. $t_z^{BS}\sim e^{-d_{\perp}/\xi}$, where $\xi$ characterizes the decaying length. Consequently, the effective hopping integral between adjacent superconducting bilayers follows $t_z^{BB}\sim (t_z^{BS})^2$, and the IJC follows $J_{\perp}^{BB}\propto (t_z^{BB})^2\sim (t_z^{BS})^4\sim e^{-4d_{\perp}/\xi}$. 

In quasi-2D layered superconductors such as La$_5$Ni$_3$O$_{11}$ and La$_3$Ni$_2$O$_{7}$, the bulk $T_c$ is related to the IJC through the following relation in the limit of vanishingly weak IJC ($\eta\to 0$)~\cite{kopec2000superconducting},
\begin{align}
T_c\approx\rho_0\frac{\pi}{\mathrm{ln}(32/\eta)}.
\label{eq3}
\end{align}
Here $\rho_0$ indicates the intralayer phase stiffness reflecting the pairing within the bilayer subsystem, which is approximated as the pairing temperature obtained from the RPA calculation, i.e. $\omega_De^{-1/\lambda}$. The anisotropy parameter $\eta$ reflects the strength of the tiny IJC, i.e. $\eta\sim J_{\perp}^{BB}$. Since 
$J_{\perp}^{BB}\propto (t_z^{BS})^4$, and $t_z^{BS}\sim 10^{-2}$, we have $\eta\sim 10^{-8}$ for La$_5$Ni$_3$O$_{11}$, which is extremely weak. In contrast, for La$_3$Ni$_2$O$_{7}$, since there exists no intervening normal-metallic single-layer to go across, we have $\eta\sim 10^{-4}$. Substituting this into Eq.~\eqref{eq3} yields an estimated $T_c$ for La$_5$Ni$_3$O$_{11}$. As shown in Fig.~\ref{fig6}(c), for La$_5$Ni$_3$O$_{11}$, the anisotropy parameter $\eta$ locates within the regime where a slight enhancement of $\eta$ will prominently enhance the $T_c$; while for La$_3$Ni$_2$O$_{7}$, $\eta$ has only a minor effect on $T_c$. It should be noted that the $T_c$ here refers to the three-dimensional bulk superconducting $T_c$, which exhibits true zero resistance, unlike the two-dimensional Kosterlitz–Thouless transition temperature $T_{KT}$, where the resistance (or conductance) exhibits a power-law decay to zero with voltage.

Under applied pressure, the strain is proportional to the pressure, implying that $\Delta d_{\perp}\propto-P$, i.e., $d_{\perp}(P)=d_{\perp}^{0}-\kappa P$, where $d_{\perp}^{0}$ is the interlayer distance at zero pressure and $\kappa$ is the modulus constant. Therefore,
\begin{align}
\eta\sim e^{-4(d_{\perp}^{0}-\kappa P)/\xi}\equiv \eta_0e^{\alpha P}. 
\end{align}
Here, $\eta_0$ is the anisotropy parameter at AP. Ultimately, 
\begin{align}
T_c\approx T_c^{\mathrm{RPA}}\frac{\pi}{\mathrm{ln}(\frac{32}{\eta})}\approx \omega_De^{-1/\lambda}\frac{\pi}{\mathrm{ln}(\frac{32}{\eta_0e^{\alpha P}})}. 
\end{align}
Since $t_z^{BS}$ is too small for DFT to capture its pressure dependence with sufficient accuracy, a precise calculation is challenging; therefore, we extract the parameters by fitting to experiment. For a reasonable set of parameters, e.g. $\eta_0\approx 10^{-8}$ and $\alpha \approx0.3$ GPa$^{-1}$, the resulting pressure dependence of $T_c$ shown in Fig.~\ref{fig6}(d) indeed exhibits a dome-shaped behavior similar to that observed in experiment~\cite{shi2025superconductivityhybridruddlesdenpopperla5ni3o11}. 





~~~~~~~~~~~~~~~~~~~~~~~~~~~~~~~~~~~~~~~~~~~~~~~~~~~~~~~~~~~~~~~~~~~~~~~~~~~~~~~~~~~~

\noindent{\bf Discussion}

\noindent In this work, we investigate the electronic structure and superconducting mechanism of the hybrid RP nickelate La$_5$Ni$_3$O$_{11}$, composed of alternating single-layer and bilayer subsystems. Based on DFT and RPA calculations, we find that SC predominantly emerges in the bilayer subsystem with an $s^{\pm}$-wave pairing symmetry, closely resembling that in pressurized bulk La$_3$Ni$_2$O$_7$. We further show that the pressure dependence of the superconducting $T_c$ in La$_5$Ni$_3$O$_{11}$ is governed not only by the intralayer pairing but also by the IJC, which leads to a dome-like $T_c(P)$ behavior consistent with experimental observations. Although our analysis reproduces this dome-like trend, this model should be regarded as one possible channel to account for the observed $T_c$ of La$_5$Ni$_3$O$_{11}$, while other channels may also be relevant.

\begin{figure}[htbp]
\centering
\includegraphics[width=0.35\textwidth]{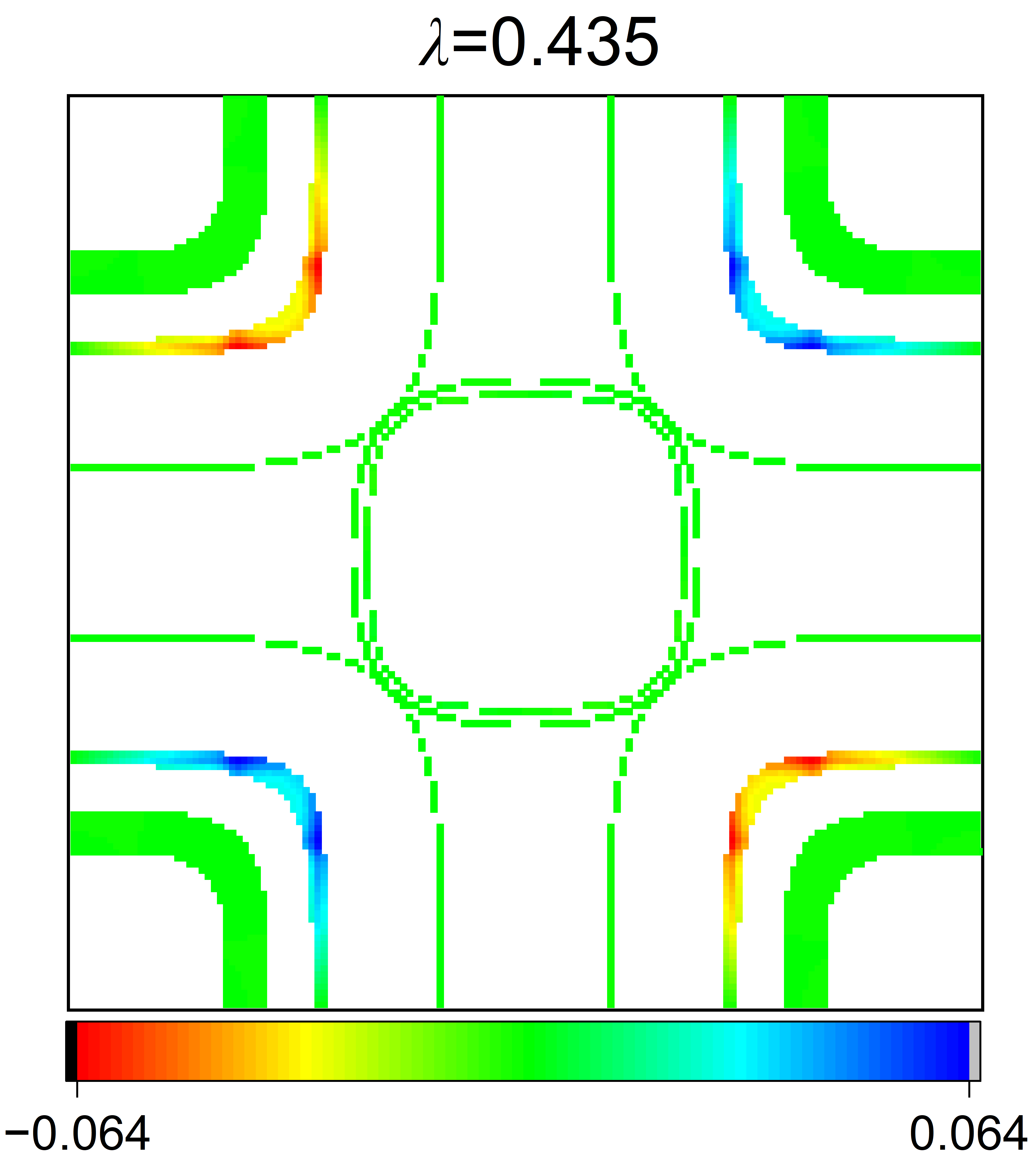}
\caption{(color online) The distributions of the leading $d_{xy}$-wave pairing gap function with $\lambda=0.435$ at $U=0.5$ eV, $J_H=1/6U$.}
\label{fig7}
\end{figure}

{\color{red}Recently, a full monolayer–bilayer RPA study was reported in Ref.~\cite{zhang2025electronicstructuremagneticpairing}. We therefore also performed a full monolayer–bilayer calculation, as shown in Fig.~\ref{fig7}.}

~~~~~~~~~~~~~~~~~~~~~~~~~~~~~~~~~~~~~~~~~~~~~~~~~~~~~~~~~~~~~~~~~~~~~~~~~~~~~~~~~~~~~~~~~~~~~~~~~~~~~~~~~~~~~~~~~~~~~~~~~~~~~~~~~~~~~~~~~~~~~~~

\noindent{\bf Methods}

\noindent{\bf DFT method}

\noindent DFT calculations were performed by Vienna ab initio simulation package (VASP)~\citep{VASP1,VASP2}, in which the projector augmented wave (PAW)~\citep{PAW1,PAW2} method with a 600 eV plane-wave cutoff is applied. 
The generalized gradient approximation (GGA) of Perdew-Burke-Ernzerhof (PBE) form exchange correlation potential is adopted ~\citep{PhysRevLett.77.3865}.
 The convergence criterion of force was set to be 0.001\ eV/{\rm \AA}  and total energy convergence criterion was set to be $10^{-7}$ eV. A $\Gamma$-centered $19\times19\times5$ Monkhorst Pack k-mesh grid is used for primitive cell of $P4/mmm$ phase and a $\Gamma$-centered $14\times14\times5$ Monkhorst Pack k-mesh grid for primitive cell of $Cmmm$ phase.
In structural relaxations, the lattice constants of $P4/mmmm$ phase we use are $a = 3.77$ \AA\  and $c = 16.20$ \AA. DFT calculations reveal that the corresponding pressure is 12 GPa, where La$_5$Ni$_3$O$_{11}$ enters superconducting state ~\cite{shi2025superconductivityhybridruddlesdenpopperla5ni3o11}. An effective Hubbard U parameter $U=3.5$ eV was employed to account for the correlation effects of 3d electrons in Ni atoms~\citep{yang_orbital-dependent_2024,PhysRevB.57.1505}. 
To obtain the projected TB models, we further performed Wannier downfolding  as implemented in WANNIER90~\citep{w90} package, in which  the good convergences were reached.

~~~~~~~~~~~~~~~~~~~~~~~~~~~~~~~~~~~~~~~~~~~~~~~~~~~~~~~~~~~~~~~~~~~~~~~~~~~~~~~~~~~~~~~~~~~~~~~~~~~~~~~~~~~~~~~~~~~~~~~~~~~~~~~~~~~~~~~~~~~~~~~

\noindent{\bf RPA method}

\noindent In the standard multi-orbital RPA approach, the bare susceptibility is defined as
\begin{align}\label{chi01}
\chi^{(0)pq}_{st}(\bm{k},\tau)\equiv
&\frac{1}{N}\sum_{\bm{k}_1\bm{k}_2}\left\langle
T_{\tau}c_{p}^{\dagger}(\bm{k}_1,\tau)
c_{q}(\bm{k}_1+\bm{k},\tau)\right.                      \nonumber\\
&\left.\times c_{s}^{\dagger}(\bm{k}_2+\bm{k},0)
c_{t}(\bm{k}_2,0)\right\rangle_0,
\end{align}
where $\langle\cdots\rangle_0$ represents the expectation value in the free-electron state, and $p/q/s/t$ are the effective orbital indices, which label combined layer, sublattice and physical orbital indices. Transforming the above defining formula to the momentum-frequency space, we obtain the explicit formula of bare susceptibility as
\begin{align}\label{chi0e}
\chi^{(0)pq}_{st}&(\bm{k},i\omega_n)
=\frac{1}{N}\sum_{\bm{k}_1\alpha\beta}
\xi^{*}_{\alpha p}(\bm{k}_1)
\xi_{\beta q}(\bm{k}_1+\bm{k})
\xi^{*}_{\beta s}(\bm{k}_1+\bm{k})                        \nonumber\\
&\xi_{\alpha t}(\bm{k}_1)
\frac{f(\varepsilon^{\beta}_{\bm{k}_1+\bm{k}}-\mu_c)
-f(\varepsilon^{\alpha}_{\bm{k}_1}-\mu_c)}
{i\omega_n+\varepsilon^{\alpha}_{\bm{k}_1}
-\varepsilon^{\beta}_{\bm{k}_1+\bm{k}}},
\end{align}
where $\alpha,\beta$ represent band indices, $\varepsilon$ and $\xi$ are the eigen-value, and eigen-state of the free particle Hamiltonian, $\mu_c$ is the chemical potential, and $f(\varepsilon_{\bm{k}})=1/(1+e^{\beta \varepsilon_{\bm{k}}})$ is the Fermi-Dirac function.

In the RPA level,  the renormalized spin/charge susceptibilities for the system are
\begin{align}\label{chisce1}
 \chi^{(s)}(\mathbf {q},i\nu)=[I-\chi^{(0)}(\mathbf {q},i\nu)
 U^{(s)}]^{-1}\chi^{(0)}(\mathbf {q},i\nu),\nonumber\\
 \chi^{(c)}(\mathbf {q},i\nu)=[I+\chi^{(0)}(\mathbf {q},i\nu)
 U^{(c)}]^{-1}\chi^{(0)}(\mathbf {q},i\nu).
\end{align}
Where $\chi^{(s,c)}(\mathbf {q},i\nu)$, $\chi^{(0)}(\mathbf {q},i\nu)$, and $U^{(s,c)}$ are operated as $l^2$ $\times$ $l^2$ matrices ($l$ represents the number of orbits and the upper or lower two indices are viewed as one
number) with elements of the matrix $U^{(s/c)}$ to be
$$ U^{(s)pq}_{st}=\left\{
\begin{aligned}
U , ~~~p=q=s=t \\
J_H,~~~ p=q\neq s=t \\
J_H,~~~ p=s\neq q=t \\
V,~~~ p=t\neq s=q\\
\end{aligned}
\right.
$$

$$ U^{(c)pq}_{st}=\left\{
\begin{aligned}
U , ~~~p=q=s=t \\
2V-J_H,~~~ p=q\neq s=t \\
J_H,~~~ p=s\neq q=t \\
2J_H-V,~~~ p=t\neq s=q\\
\end{aligned}
\right.
$$
Since we only consider the on-site interaction, the elements of the matrix $U^{(s/c)}$ are non-zero only if $p,q,s,t$ are the same layer indices.

~~~~~~~~~~~~~~~~~~~~~~~~~~~~~~~~~~~~~~~~~~~~~~~~~~~~~~~~~~~~~~~~~~~~~~~~~~~~~~~~~~~~~~~~~~~~~~~~~~~~~~~~~~~~~~~~~~~~~~~~~~~~~~~~~~~~~~~~~~~~~~~

Note that there is a critical interaction strength $U_c^{(s,c)}$ for spin and charge respectively. When $U>U_{c}^{(s,c)}$,  the denominator matrix in Eq.~\ref{chisce1} will have zero eigenvalues for some $\mathrm{q}$ and the renormalized spin
or charge susceptibility diverges there, which invalidates the RPA treatment~\cite{takimoto2004strong,yada2005origin,kubo2007pairing,kuroki101unconventional,graser2009near,liu2013d+,RPA7,RPA8,PhysRevB.110.L180501}.  This divergence of spin susceptibility
for $U>U_{c}^{(s)}$ implies magnetic order, while that of the charge susceptibility for $U>U_{c}^{(c)}$ implies charge order.

~~~~~~~~~~~~~~~~~~~~~~~~~~~~~~~~~~~~~~~~~~~~~~~~~~~~~~~~~~~~~~~~~~~~~~~~~~~~~~~~~~~~~~~~~~~~~~~~~~~~~~~~~~~~~~~~~~~~~~~~~~~~~~~~~~~~~~~~~~~~~~~

When $U<U_c$, a Cooper pair $c_{t}(\mathbf {q})c_{s}(-\mathbf {q})$ could be scattered to $c_{p}^{\dagger}(\mathbf {k})c_{q}^{\dagger}(-\mathbf {k})$ by exchanging charge or spin fluctuations. Considering only intra-band pairings, we obtain the following effective pairing interaction:
\begin{align}\label{veff}
V_{eff}=\frac{1}{N}\sum_{\alpha\beta\mathbf {k}\mathbf{q}}V^{\alpha\beta}(\mathbf {k},\mathbf{q})c_{\alpha}^{\dagger}(\mathbf {k})c_{\alpha}^{\dagger}(-\mathbf {k})c_{\beta}(-\mathbf{q})c_{\beta}(\mathbf{q})
\end{align}
Here $\alpha$/$\beta$ is band indice. And effective pairing interaction vertex $V^{\alpha\beta}(\mathbf {k},\mathbf{q})$ has the form:
\begin{align}\label{v}
V^{\alpha\beta}(\mathbf {k},\mathbf {q})=\sum_{pqst}\Gamma _{pq}^{st}(\mathbf {k},\mathbf {q})\xi^{\alpha,*}_{p}(\mathbf {k})\xi_{q}^{\alpha,*}(-\mathbf {k})\xi_{s}^{\beta}(-\mathbf {q})\xi_{t}^{\beta}(\mathbf {q}).
\end{align}
Here $\xi^{\alpha}\left(\mathbf {k}\right)$ are the  eigenvector corresponding to $\alpha$-th eigenvalue (relative to the chemical potential $\mu_c$) of the TB Hamiltonian matrices. Within the mean-field approximation, one can derive a self-consistent equation for the pairing gap, which when linearized near $T_c$~\cite{takimoto2004strong,yada2005origin,kubo2007pairing,kuroki101unconventional,graser2009near,liu2013d+,RPA7,RPA8,PhysRevB.110.L180501} yields the linearized gap equation~(\ref{gapfull}):
\begin{align}\label{gapfull}
\Delta_{\alpha}(\mathbf {k})=-\sum_{\beta\mathbf{q}}V^{\alpha\beta}(\mathbf {k},\mathbf{q})\times \frac{\tanh(\frac{\beta_c}{2}\left 
 |\tilde{\varepsilon}_{\beta}(\mathbf{q})\right |)}{\left |\tilde{\varepsilon}_{\beta}(\mathbf{q})\right |}\times\Delta_{\beta}(\mathbf{q}).
\end{align}
Here $\beta_c$=$\frac{1}{\mathrm{k_B}T_c}$ represents the critical temperature of SC. Choosing a thin energy shell near the Fermi level, Eq.~(\ref{gapfull}) becomes the eigenvalue problem of the effective interaction matrix $V^{\alpha\beta}$, which  determines the $T_c$ and the leading pairing symmetry of the system. To be specific, $\Delta_{\alpha}(\mathbf {k})$ represents the relative gap function on the $\alpha$-th FS patches near $T_c$, and eigenvalue $\lambda$ is related to $T_c$ through $\lambda^{-1}=\mathrm{ln}(1.13\frac{\hbar\omega_D}{k_{\mathrm{B}}}T_c)$. The leading pairing symmetry is determined by the largest eigenvalue $\lambda$ of Eq.~(\ref{gap}):
\begin{equation}\label{gap}
-\frac{1}{(2\pi)^2}\sum_{\beta}\oint_{FS}
dq_{\Vert}\frac{V^{\alpha\beta}(\mathbf {k},\mathbf{q})}{v^{\beta}_{F}(\mathbf{q})}\Delta_{\beta}(\mathbf{q})=\lambda
\Delta_{\alpha}(\mathbf {k}).
\end{equation}	

~~~~~~~~~~~~~~~~~~~~~~~~~~~~~~~~~~~~~~~~~~~~~~~~~~~~~~~~~~~~~~~~~~~~~~~~~~~~~~~~~~~~~~~~~~~~~~~~~~~~~~~~

\noindent{\bf Data availability}

\noindent Relevant data supporting the key findings of this study are available within the article and the Supplementary Information file. All raw data generated during the current study are available from the corresponding authors upon reasonable request.

~~~~~~~~~~~~~~~~~~~~~~~~~~~~~~~~~~~~~~~~~~~~~~~~~~~~~~~~~~~~~~~~~~~~~~~~~~~~~~~~~~~~~~~~~~~~~~~

\noindent{\bf Code availability}

\noindent The code that supports the plots within this paper is available from the corresponding author upon reasonable request.

~~~~~~~~~~~~~~~~~~~~~~~~~~~~~~~~~~~~~~~~~~~~~~~~~~~~~~~~~~~~~~~~~~~~~~~~~~~~~~~~~~~~~~~~~~~~~~~~~~~~~~~~~~~~~~~~~~~~~~~~~~~~~~~~~~~~~~~~~~~~~~~

\noindent{\bf References}
\bibliographystyle{naturemag.bst}
\bibliography{references}

~~~~~~~~~~~~~~~~~~~~~~~~~~~~~~~~~~~~~~~~~~~~~~~~~~~~~~~~~~~~~~~~~~~~~~~~~~~~~~~~~~~~~~~~~~~~~~~~~~~~~~~~~~~~~~~~~~~~~~~~~~~~~~~~~~~~~~~~~~~~~~~

\noindent{\bf Acknowledgements}

\noindent We are grateful for the discussion with Chen Lu.  This work is supported by National Natural Science Foundation of China (Grants No. 12494591, No. 92165204, No. 12234016, No. 12074031), National Key Research and Development Program of China (2022YFA1402802), Guangdong Provincial Key Laboratory of Magnetoelectric Physics and Devices (2022B1212010008), Guangdong Fundamental Research Center for Magnetoelectric Physics (2024B0303390001), and Guangdong Provincial Quantum Science Strategic Initiative (GDZX2401010). Ming Zhang is supported by the Zhejiang Provincial Natural Science Foundation of China under Grant No. ZCLQN25A0402.

~~~~~~~~~~~~~~~~~~~~~~~~~~~~~~~~~~~~~~~~~~~~~~~~~~~~~~~~~~~~~~~~~~~~~~~~~~~~~~~~~~~~~~~~~~~~~~~~~~~~~~~~~~~~~~~~~~~~~~~~~~~~~~~~~~~~~~~~~~~~~~~

\noindent{\bf Author contribution}

\noindent Dao-Xin Yao and Fan Yang conceived and designed the project.  Ming Zhang  and Cui-Qun Chen performed the numerical calculations for DFT, TB model, and RPA calculations. All authors contributed to the discussion of the results and wrote the paper.

~~~~~~~~~~~~~~~~~~~~~~~~~~~~~~~~~~~~~~~~~~~~~~~~~~~~~~~~~~~~~~~~~~~~~~~~~~~~~~~~~~~~~~~~~~~~~~~~~~~~~~~~~~~~~~~~~~~~~~~~~~~~~~~~~~~~~~~~~~~~~~~

\noindent{\bf Competing interests}

\noindent The authors declare no competing interests.

~~~~~~~~~~~~~~~~~~~~~~~~~~~~~~~~~~~~~~~~~~~~~~~~~~~~~~~~~~~~~~~~~~~~~~~~~~~~~~~~~~~~~~~~~~~~~~~~~~~~~~~~~~~~~~~~~~~~~~~~~~~~~~~~~~~~~~~~~~~~~~~

\noindent{\bf Addition information}

\noindent{\bf Supplementary information} The online version contains supplementary information available at
\end{document}